\documentclass[%
 reprint,
 superscriptaddress,
 amsmath,
 amssymb,
 aps,
 pra,
 twocolumn,
 floatfix,
 tightenlines
]{revtex4-2}

\usepackage{graphicx}

\begin{document}

\title{Feasibility considerations for free-fall tests of gravitational decoherence} 

\author{R. Kaltenbaek}
 \email{rainer.kaltenbaek@fmf.uni-lj.si}
 \affiliation{Faculty of Mathematics and Physics, University of Ljubljana, Ljubljana, Slovenia}
 \affiliation{Institute for Quantum Optics and Quantum Information - Vienna, Vienna, Austria}

\begin{abstract}
\noindent Space offers exciting opportunities to test the foundations of quantum physics using macroscopic quantum superpositions. It has been proposed to perform such tests in a dedicated space mission (MAQRO) using matter-wave interferometry with massive test particles or monitoring how the wave function of a test particle expands over time. Such experiments could, test quantum physics with sufficiently high precision to resolve potential deviations from its unitary evolution due to gravitational decoherence. For example, such deviations have been predicted by the Di\'{o}si-Penrose (DP) model and the K\'{a}rolyh\'{a}zy (K) model. The former predicts the collapse of massive or large superpositions due to a non-linear modification of quantum evolution. The latter predicts decoherence because of an underlying uncertainty of space time. Potential advantages of a space environment are (1) long free-fall times, (2) low noise, and (3) taking a high number of data points over several years in a dedicated space mission. In contrast to interferometric tests, monitoring wave function expansion is less complex, but it does face some practical limitations. Here, we will discuss limitations of such non-interferometric experiments due to the limited number of data points achievable during a mission lifetime. Our results show that it will require an interferometric approach to conclusively test for gravitational decoherence as predicted by the DP or K models. In honor of the novel prize of Sir Roger Penrose, we will focus our discussion on the  Di\'{o}si-Penrose model.
\end{abstract}


\maketitle

\section{Introduction}
It has long been contended whether the predictions of quantum physics still apply for increasingly macroscopic objects or whether a transition will occur to our everyday ``classical'' world. Schr\"{o}dinger highlighted this issue with his famous cat\cite{Schroedinger1935a}. Still, the predictions of quantum physics continue being confirmed in matter-wave experiments with increasingly massive test particles. Currently the mass record are molecules consisting of $\sim 10^3$ atoms\cite{Fein2019a}. Already in the 1960s, K\'{a}rolyh\'{a}zy predicted a natural limit to macroscopic superpositions due to decoherence caused by an inherent uncertainty of space time\cite{Karolyhazy1966a}. Decoherence due to quantum gravity was also predicted by Ellis, Mohanty and Nanopoulos~\cite{Ellis1984a,Ellis1989a,Ellis1992a}. Collapse models that introduce small nonlinear modifications to quantum time evolution have claimed to solve the quantum measurement problem. The nonlinear modifications aim at an objective collapse of macroscopic wave functions\cite{Bassi2013a}. One of the most well known collapse model is the one by Ghirardi, Rimini and Weber\cite{Ghirardi1986a} (GRW). A generalization of that model is is the continuous-spontaneous-localization (CSL) model\cite{Ghirardi1990a}. Di\'{o}si proposed that gravity could be the cause of such a nonlinear modification of quantum physics\cite{Diosi1984a,Diosi1987a}. Penrose suggested\cite{Penrose1998a,Penrose1996a,Penrose1998b,Penrose2014a} that macroscopic superpositions should have a limited lifetime because they lead to a superposition of distinct space times. He proposed that such superposition states would decay exponentially with a life-time inversely proportional to the gravitational self-energy of the wave function. Because the models of Di'{o}si and Penrose lead to the same predictions\cite{Diosi2005a,Diosi2007a}, we will refer to them as the Di'{o}si-Penrose (DP) model. The benefits of the DP model in its most general form is that it does not depend on any parameters. While more restrictive versions of the DP model have already been ruled out\cite{Donadi2021a}, experiments proposing to test this general form of the DP model are more challenging. For example, it was suggested to test the DP model using dielectric particles in space\cite{Kaltenbaek2012b,Kaltenbaek2016a,Voirin2019a}, or by using superconducting spheres in ground-based experiments\cite{Pino2018a}. In either of those experiments we might expect seeing deviations from quantum physics due to gravitational effects. 

\section{Free-fall tests of quantum physics}
\label{sec::freeFallTests}
In 2010, it was proposed to use microgravity in space for matter-wave experiments to test quantum physics with high-mass dielectric test particles\cite{Kaltenbaek2012b}. From the outset, a central goal of this mission proposal (MAQRO) was to see whether we could test quantum physics in a parameter regime where one might expect deviations from gravity. For that purpose, we have been using the model of Di\'{o}si and Penrose as a central benchmark.

We later updated the MAQRO mission proposal\cite{Kaltenbaek2016a} to use near-field matter-wave interferometry to allow for higher-visibility interference, and for shorter free-fall times ($\le 100\,$s compared to several $100\,$s in the original proposal). Another important addition to the updated MAQRO proposal was to also perform non-interferometric tests. The central idea was the following:
\begin{enumerate}
    \item trap a test particle and cool its center-of-mass (CM) motion close to the ground state
    \item release the particle and let the wave function expand for some time $t$
    \item measure the position of the test particle
    \item repeat this procedure many times
\end{enumerate}

By looking at the distribution of measured positions for at least two times, one can estimate the width of the wave function at different times. We can then compare how quickly the wave function expands with the predictions of quantum physics and with the predictions of alternate theoretical models.

The core advantages of this non-interferometric approach are that (1) it is less complex than an interferometric test, (2) in contrast to interferometric tests, there is no maximum amount of decoherence this approach can deal with. In interferometric tests, if the decoherence is too strong, we will simply not see an interference pattern, and it may be hard to find out why that is the case. The latter point was the main reason for including this approach in MAQRO. The intention was that these measurements could bridge a potential gap between ground-based and space-based experiments. Also, this would provide a second method to measure decoherence and in this way calibrate the results one then would get with the interferometric approach\cite{Kaltenbaek2016a}.

When considering such free-fall experiments to measure the expansion of the wave function, it quickly became clear that it will be even more crucial to take into account measurement statistics than it would be in ground-based experiments. One of the main advantages of space experiments are the long free-fall times. At the same time, any space mission will have a limited lifetime. Even if the life time is a few years, if the time per data point is $100\,$s, this will pose a noticeable upper limit on the measurement statistics achievable\cite{Kaltenbaek2016a}. We refined these considerations in the proposal of MAQRO for a call of the European Space Agency (ESA) for New Science Ideas, and these considerations were later published in the report of a feasibility study based on MAQRO\cite{Voirin2019a}. Here, we will present a comprehensive overview of these arguments, we will include new considerations and the potential for future improvements. In particular, we will focus on the implications for potential future tests of the DP model.

\section{Practical considerations and statistical uncertainty}
\label{sec::practical}
Space missions typically have a limited life time for various reasons. To name a few, this could be budget constraints, limited fuel supply, limited supply of cooling agents, or material degradation in a space environment. If there are multiple measurement series, this will cause additional constraints. For example, the tests of quantum physics we consider here would be performed with particles of varying radii and varying mass density. That will limit the time for one measurement series to, at most, a tenth of the mission life time. In addition, some time will be needed for station keeping maneuvers, for calibration measurements, and there should be some time left for error analysis and mitigation. Here, we will assume a maximum measurement time $T$ of \textbf{30 days} (1 month) per measurement series.

If we want to observe the dependence of wave-function expansion on time, we need to either measure the width of the wave function at least at two different times, or one prepares the initial state very well and then measures the width of the wave function once at a later time $t$. For simplicity, we will assume the later approach. It is also the approach requiring the least amount of time. According to quantum physics, if we assume an evolution in 1D, if the wave function is centered at the coordinate origin, and if our particle's mean velocity is zero, the variance of the wave function will be
\begin{equation} 
    \langle \hat{x}^2(t)\rangle = \langle \hat{x}^2(0)\rangle + \frac{t^2}{m^2}\langle \hat{p}^2(0)\rangle + \frac{2\Lambda \hbar^2}{3m^3}t^3.\label{eq::variance}
\end{equation}
$\hat{x},\hat{p}$ are the position and momentum operators, respectively, of the center-of-mass of our test particle. The last term in this equation is proportional to a decoherence parameter $\Lambda$. For $\Lambda=0$, we have pure unitary evolution according to Schr\"{o}dinger's equation. Non-vanishing values of $\Lambda$ can be used to describe standard decoherence effects in the long-wavelength limit\cite{Schlosshauer2007a} or possible deviations from the predictions of quantum physics\cite{Kaltenbaek2012b,RomeroIsart2011c}.

The main goal of the experimental tests we want to describe here will be to determine the value of $\Lambda$ and its dependence on changes to the properties of the test particles used (radius, mass density). This parameter dependence will be different for various decoherence mechanisms and for alternate theoretical models. The precision with which we can determine $\Lambda$ will be determined by the precision with which we can determine $\langle \hat{x}^2(t)\rangle$. If we assume that we know $t$, $m$, and the variance of the initial state $\langle \hat{x}^2(0)\rangle$ to sufficient accuracy, the main inaccuracies will be (1) the measurement accuracy, and (2) the statistical uncertainty.

In a recent paper, a sufficiently high measurement accuracy was achieved for feedback cooling optically optically trapped particles close to the quantum ground state. The achieved readout accuracy was better than $0.1\,$pm for the trap frequencies used. This accuracy will be reduced after long expansion times because the waist of the optical readout beams will have to be correspondingly larger. Instead of the $1\,\mu$m waist in this experiment, we will require a waist of about $1\,$mm to ensure that the particle will still be within the optical beam used for the measurement\cite{Voirin2019a}. This will reduce the measurement accuracy by approximately $6$ orders of magnitude to better than $100\,$nm. Let us estimate how that measurement inaccuracy will compare with the statistical uncertainty.

The fractional statistical uncertainty with which we can determine the variance $\langle\Delta x^2\rangle$ of the wave packet is\cite{Taylor1997a}:
\begin{equation}
    \frac{\delta\left(\Delta x^2\right)}{\langle\Delta x^2\rangle} = \sqrt{\frac{2}{\mathcal{N}-1}} = \sqrt{\frac{2 t}{T-t}}\approx \sqrt{\frac{2 t}{T}}.
\end{equation}
Here, $\mathcal{N}$ is the number of measurements, $\langle\cdot\rangle$ denotes the ensemble average, and $t$ is the time for a single measurement. The approximation assumes $T\gg t$.

In order to see deviations from the predictions of quantum physics, we will want to reduce any unwanted decoherence effects that could mask those deviations. Under these conditions, we can expect the modifications proportional to $\Lambda$ in the third term of equation \ref{eq::variance} to be small compared to the first two parts in that equation. For long expansion times $t$, we can also assume that the second term dominates over the first. That means, we can write:
\begin{equation}
    \frac{\delta\left(\Delta x^2\right)}{\langle\Delta x^2\rangle} \approx \frac{\delta\left(\Delta x^2\right)}{\langle \hat{x}^2(0)\rangle + \frac{t^2}{m^2}\langle \hat{p}^2(0)\rangle}\approx \frac{\delta\left(\Delta x^2\right)}{\frac{t^2}{m^2}\langle \hat{p}^2(0)\rangle}.\label{eq::statisticalUncertainty}
\end{equation}

Even for a comparatively large measurement uncertainty of $100\,$nm, this statistical uncertainty will already dominate over the measurement uncertainty after a very short time. If we assume a test particle mass of $10^9\,$ atomic mass units (amu), and a trap frequency of $10^5\,$Hz, and a $T$ of 30 days, the statistical uncertainty will already dominate over the measurement uncertainty for $t\gtrsim 1\,$s.

In order to study the feasibility of testing gravitational decoherence, we will therefore focus on the limitations resulting from the statistical uncertainty in space-based free-fall experiments. Given this statistical uncertainty, there will be a minimum $\Lambda$ we will be able to distinguish from pure quantum evolution:
\begin{equation}
    \Lambda_{\mathrm{min}} = \sqrt{\frac{1}{2 T t}} \frac{3 \langle \hat{p}^2(0)\rangle}{\hbar^2}. \label{eq::minLambda}
\end{equation}

It is worth noting that, for a fixed total time $T$, it is preferential to choose $t$ as large as possible as long as it stays much smaller than $T$ although that will reduce $\mathcal{N}=T/t$.

\section{The feasibility of testing for gravitational decoherence}
\label{sec::testingDP}
According to Penrose, macroscopic superpositions will decohere on a time scale $\tau_G$ defined by their gravitational self energy $E_G$\cite{Penrose1998a,Penrose1996a,Penrose1998b,Penrose2014a}:
\begin{equation}
    \tau_G \approx \frac{\hbar}{E_G}.
\end{equation}
The gravitational self-energy of a superposition of a test mass being at position $\mathbf{x}$ or at position $\mathbf{y}$ is assumed to be the gravitational potential between two mass distributions centered at $\mathbf{x}$ and $\mathbf{y}$\cite{Diosi1984a,Diosi1987a,Diosi2005a,Diosi2007a,Penrose2014a}.

The effect of this gravitational decoherence on the density matrix can be described following the description of Di\'{o}si\cite{Diosi2007a}:
\begin{widetext}
\begin{equation}
    \frac{\partial\rho(x,y,t)}{\partial t} = \frac{i\hbar}{2 m}\left(\frac{\partial^2}{\partial x^2}-\frac{\partial^2}{\partial y^2}\right)\rho(x,y,t)+\frac{U(\mathbf{x},\mathbf{x})+U(\mathbf{y},\mathbf{y})-2 U(\mathbf{x},\mathbf{y})}{2\hbar}
\end{equation}
\end{widetext}
Here, the gravitational potential between test particles at positions $\mathbf{x}$ and $\mathbf{y}$ is\cite{Diosi2007a}:
\begin{equation}
    U(\mathbf{x},\mathbf{y}) = -G \int d\mathbf{r}\int d\mathbf{r}^{\prime} \frac{f(\mathbf{r},\mathbf{x})f(\mathbf{r}^{\prime},\mathbf{y})}{\vert\mathbf{r}^{\prime}-\mathbf{r}\vert}.
\end{equation}
$f(\mathbf{r},\mathbf{x})$ is the mass density at position $\mathbf{r}$ if the test particle is located at position $\mathbf{x}$.

For example, a recent paper by Donadi et al\cite{Donadi2021a} that claims to have ruled out a ``parameter independent'' version of the DP model assumed the mass distribution to be strongly localized around the positions of atoms arranged within a lattice in the test particle. This results in a very strong gravitational self-energy, short decoherence times, and a significant amount of heating incompatible with measurement results unless one introduces a cut-off parameter.

Given that gravitation in a low-energy setting typically varies very slowly with distance, the present author finds it more reasonable to assume a continuous mass distribution to calculate the gravitational self energy. This provides a lower bound on the gravitational decoherence one can expect from this model without the need for introducing any parameters to the DP model. Assuming a continuous mass distribution also avoids the issue of too high spurious heating. In particular, one gets
\begin{equation}
    \frac{d\langle \hat{H}\rangle}{dt} = \frac{1}{2 m}\frac{d\langle \hat{p}^2\rangle}{dt} = \frac{m \hbar G}{2 a^3}
\end{equation}
for a particle of radius $a$ and mass $m$. For a fused silica particle with $a=200\,$nm, this leads to a heating rate of $10^{-18}\,\mathrm{K/s}$ compared to the $10^{-4}\,\mathrm{K/s}$ reported by Donadi et al\cite{Donadi2021a}.

For this derivation we used a definition given by Penrose\cite{Penrose2014a} for the gravitational self energy of a spherical test particle of radius $a$ with a uniform mass distribution in a superposition of size $b=\vert\mathbf{y}-\mathbf{x}\vert$:
\begin{equation}
    E_G=\left\{
    \begin{array}{ll}
         \frac{m^2 G}{a}\left(2\lambda^2-\frac{3}{2}\lambda^3+\frac{1}{5}\lambda^5\right)& \mathrm{\;for\;}0\le\lambda\le 1  \\
         \frac{m^2 G}{a}\left(\frac{6}{5}-\frac{1}{2\lambda}\right)& \mathrm{\;for\;}1\le\lambda
    \end{array}
    \right.
    \label{eq::gravSelfEnergy}
\end{equation}

The lower heating rate resulting from a continuous mass distribution, however, also makes it more difficult to see deviations from the predictions of quantum physics if one monitors the variance of an expanding wave packet. The excessive momentum diffusion according to the DP model can then be characterized by equation \ref{eq::variance} with\cite{RomeroIsart2011c}:
\begin{equation}
    \Lambda_{\mathrm{DP}} = \frac{G m^2}{2 a^3 \hbar}.
\end{equation}
The value of $\Lambda_{\mathrm{DP}}$ is extremely small, and it will be a challenge to achieve an even smaller $\Lambda_{\mathrm{min}}$. Options to achieve that can be to use high mass densities like, e.g., by using magnetically levitated superconducting spheres\cite{Pino2018a}, by using optomechanical squeezing of the initial momentum uncertainty\cite{Branford2019a}, or by using very long free evolution times as suggested in MAQRO\cite{Kaltenbaek2016a,Voirin2019a}.

\begin{figure}
\begin{center}
\includegraphics[width=0.95\linewidth]{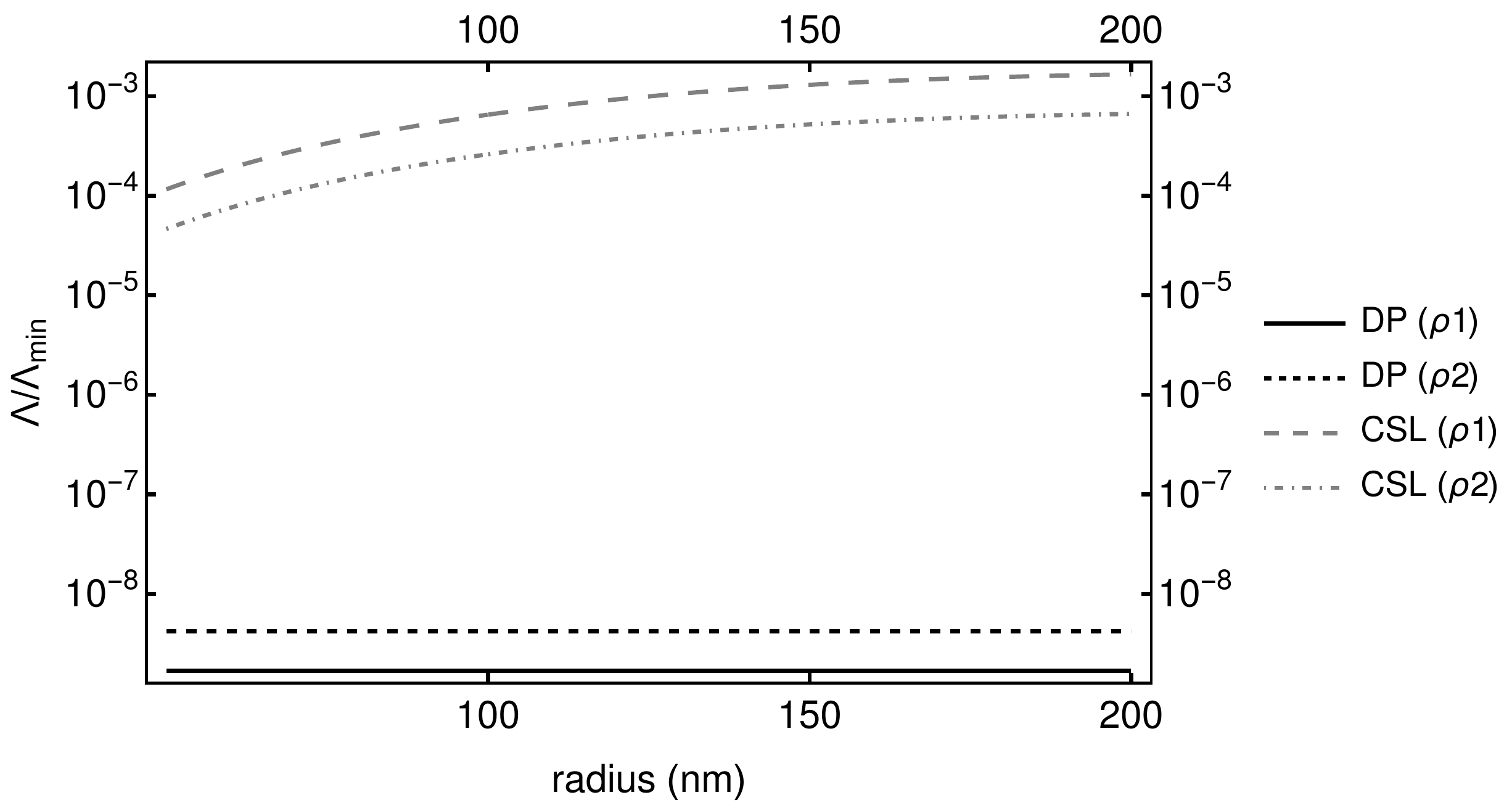}
\caption{\label{fig::expansionTime}Comparing the decoherence parameter $\Lambda$ predicted by the DP model and by the CSL model for different mass densities $\rho_1=2000\,\mathrm{kg m^{-3}}$ and $\rho_2=5000\,\mathrm{kg m^{-3}}$. We plot $\Lambda$ in units of $\Lambda_{\mathrm{min}}$ as a function of the particle radius and for a free expansion time of $t=100\,$s and a total time of 30 days for a measurement series.}
\end{center}
\end{figure}

Here, we will focus on observing free evolution, potentially enhanced by momentum squeezing. Under which conditions would it be possible to see the small increase in wave-function spreading caused by gravitational decoherence? Fig.~\ref{fig::expansionTime} shows the decoherence parameter $\Lambda$ we would expect from the DP model for test particles of different mass densities. The plot also shows the prediction of the CSL model for comparison. For the CSL prediction we used the CSL parameters proposed in the original publication\cite{Ghirardi1990a}. That means, $r_c=100\,$m for the localization radius and a localization rate of $2.2\times 10^{-17}\,$Hz.

We can see that the DP model is many orders of magnitude beyond being accessible to tests using wave-packet expansion. The original CSL model is not so far out of range, but it would still require drastic improvements - e.g, significantly longer free fall times and/or significant momentum squeezing. We can also see that a larger particle radius does not yield any advantage in testing the DP model. That is because $\Lambda_{\mathrm{DP}}$ scales with the cube of the radius of a spherical test particle, and so does $\Lambda_{\mathrm{min}}$.

\begin{figure}
\begin{center}
\includegraphics[width=0.95\linewidth]{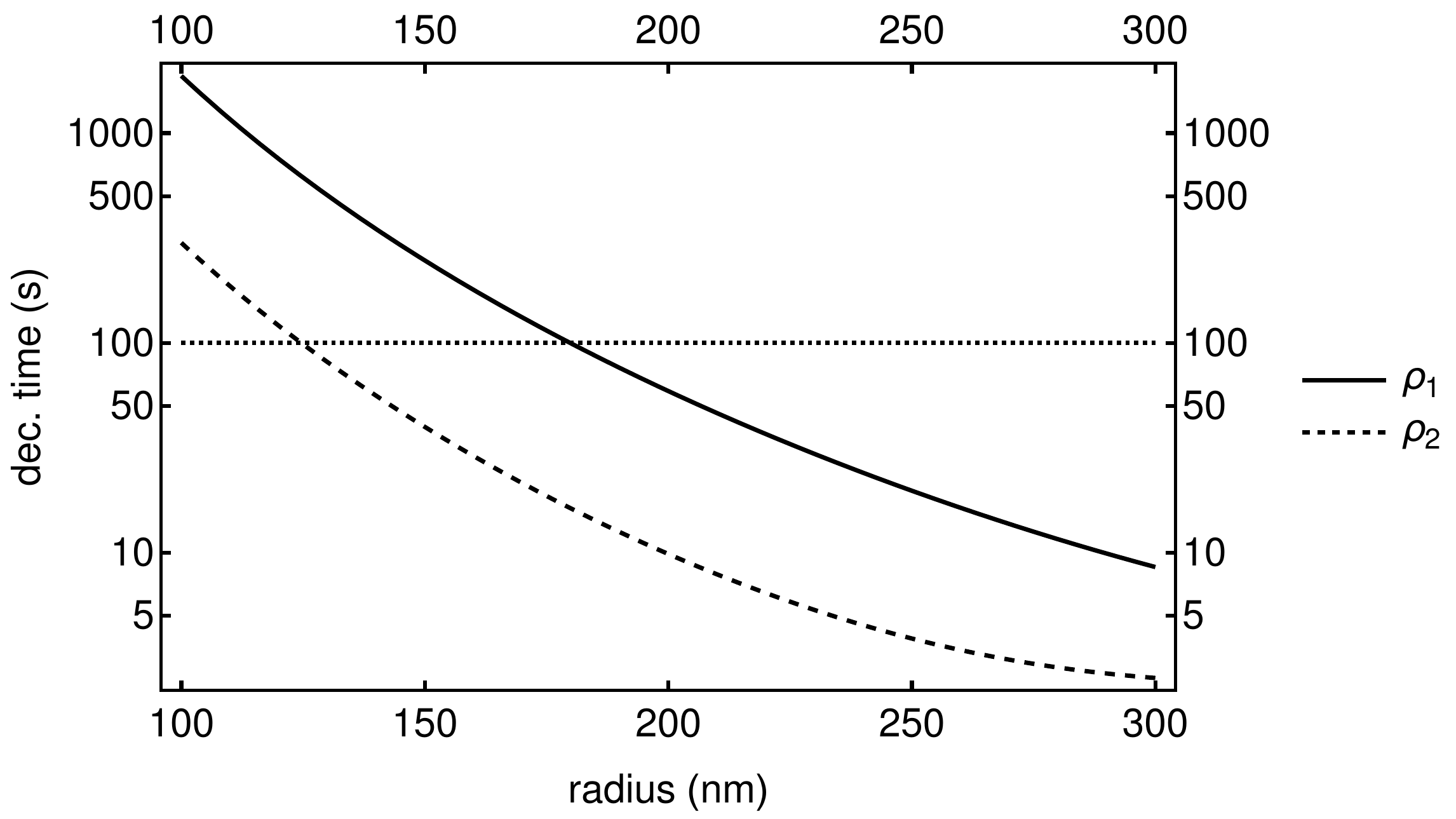}
\caption{\label{fig::decoherenceTime}Decoherence time due to gravitational decoherence. We plot the time $t_D$ after which we would expect an expanding wave packet to become significantly non-Gaussian according to the DP model. We plot the prediction for different mass densities $\rho_1=2000\,\mathrm{kg m^{-3}}$ and $\rho_2=5000\,\mathrm{kg m^{-3}}$.}
\end{center}
\end{figure}

The reason why it is so difficult to test the DP model using this approach is because the spurious heating predicted by the DP model for continuous mass densities is very small. At the same time, the decoherence predicted by the DP model can still become significant very quickly if the wave function becomes large. In order to illustrate that, let us calculate the time $t_D$ it takes a wave function to expand until the DP model predicts a decay time $t_D$. At this point, the outer parts of the wave function will start to decohere exponentially, and the wave function will become significantly non-Gaussian. We calculate this decoherence time by first defining $E_{s,G}(t)$ as the gravitational self energy for a superposition of width $b=\sqrt{\langle \hat{x}^2(t)\rangle = \langle \hat{x}^2(0)\rangle + \frac{t^2}{m^2}\langle \hat{p}^2(0)\rangle}$. This is the width of the wavepacket after a time $t$ of coherent evolution. We then solve the following equation:

\begin{equation}
    \frac{\hbar}{E_{s,G}(t)} == t.
\end{equation}

Figure \ref{fig::decoherenceTime} illustrates that it should be possible to see significant effects of the DP model already after $100\,$s for particles with a sufficiently high mass density. While this would not make a difference in wave-packet expansion, it would certainly affect matter-wave interferometry. That means, if we succeed performing matter-wave interferometry with expansion times on the order of $100\,$s with high mass densities as envisaged by MAQRO, we should be able to conclusively test the DP model. If we can, in addition, make use of momentum squeezing as proposed by Branford et al\cite{Branford2019a}, this may even become possible for shorter expansion times. This might be necessary if we cannot achieve the extremely high vacuum envisioned for MAQRO\cite{Voirin2019a}.

\section{Conclusions and Outlook}
\label{sec::conclusions}
We have discussed the observation of wave-packet expansion as a versatile tool in future free-fall experiments to test the foundations of quantum physics in space. A central challenge in such experiments will be to achieve exceedingly high particle statistics while allowing for long free evolution timed and still remaining within the limited lifetime of a space mission. In particular, we investigated whether this approach could allow tests of gravitational decoherence as predicted by a model of Di\'{o}si and Penrose (DP). We showed that conclusive tests of the DP model using wave-packet expansion seem unfeasible for the foreseeable future, but that interferometric tests as proposed in MAQRO should allow conclusive tests of this type of gravitational decoherence. A central challenge for such tests will be to achieve the vacuum conditions required for very long free expansion times on the order of $100\,$s. Moderate momentum squeezing of the initial state of the test particle might allow performing conclusive tests even for shorter free-fall times.

\begin{acknowledgments}
We are thankful for support by the Austrian Research and Promotion Agency FFG (projects 889767, 865996), and by the Slovenian Research Agency (research projects N1-0180, J2-2514, J1-9145 and P1-0125). We also want to thank M. Debiossac and L Di\'{o}si for valuable discussions.
\end{acknowledgments}

\bibliography{RKfreefall}

\end{document}